\documentclass[a4paper,fleqn,usenatbib]{mnras}

\usepackage[T1]{fontenc}
\usepackage{ae,aecompl}

\usepackage{graphicx}	
\graphicspath{{./Figures/}}
\usepackage{amsmath}	
\usepackage{amssymb}	
\usepackage{subfigmat}
\usepackage{enumerate}
\usepackage{multirow}

\title[Signatures of Protostellar Outbursts]{Observational signatures of outbursting protostars -
II: Exploring a wide range of eruptive protostars}
\author[MacFarlane  et al.]{Benjamin MacFarlane$^{1}$,
Dimitris Stamatellos$^{1}$\thanks{E-mail: dstamatellos@uclan.ac.uk},
Doug Johnstone$^{2,3}$,
Gregory Herczeg$^{4}$,
\newauthor{Giseon Baek$^{5}$,
Huei-Ru Vivien Chen$^{6}$,
Sung-Ju Kang$^{7}$,
Jeong-Eun Lee$^{5}$,
}
\\
\
\\
$^{1}$Jeremiah Horrocks Institute for Mathematics, Physics and Astronomy, University of Central Lancashire, Preston, PR1 2HE, UK\\
$^{2}$NRC Herzberg Astronomy and Astrophysics, 5071 West Saanich Rd, Victoria, BC, V9E 2E7, Canada\\
$^{3}$Department of Physics and Astronomy, University of Victoria, Victoria, BC, V8P 1A1, Canada\\
$^{4}$Kavli Institute for Astronomy and Astrophysics, Peking University, Yiheyuan 5, Haidian Qu, 100871 Beijing, China\\
$^{5}$School of Space Research and Institute of Natural Sciences, Kyung Hee University, 1732 Deogyeong-daero, Giheung-gu, \\
Yongin-si, Gyeonggi-do 446-701, Republic of Korea\\
$^{6}$Department of Physics and Institute of Astronomy, National Tsing Hua University, Taiwan\\
$^{7}$Korea Astronomy and Space Science Institute, 776 Daedeokdae-ro, Yuseong-gu, Daejeon 34055, Republic of Korea\\
}

\date{Accepted 2019 June 2. Received 2019 May 29; in original form 2019 April 27}

\pubyear{2019}

\begin{document}
\label{firstpage}
\pagerange{\pageref{firstpage}--\pageref{lastpage}}
\maketitle

\begin{abstract}

Young stars exhibit variability due to changes in the gas accretion rate onto them, an effect that should be quite significant in the early stages of their formation. As protostars are embedded within their natal cloud, this variability may only be inferred through long wavelength observations.  We perform radiative transfer simulations of young stellar objects (YSOs) formed in hydrodynamical simulations, varying the structure and luminosity properties in order to estimate the long-wavelength, sub-mm and mm, variations of their flux.  We find that the flux increase due to an outburst  event depends on the protostellar structure and is more prominent at sub-mm wavelengths than at mm wavelengths; e.g. a factor of 40 increase in the luminosity of the young protostar leads to a flux increase of  a factor of 10 at 250~\micron\ but only a factor of 2.5 at 1.3 mm. We find that the interstellar radiation field dilutes the  flux increase but that this effect may be avoided if resolution permits the monitoring of the inner regions of a YSO, where the heating is primarily due to protostellar radiation. We  also confirm that the  bolometric temperature and luminosity of outbursting protostars may result in an incorrect classification of their evolutionary stage.

\end{abstract}


%
\section{Introduction}

{Young protostars are believed to grow  in mass through episodes of intense gas accretion that result in luminosity outbursts, as seen e.g. in FU Ori-type  \citep{herbig66,herbig77} and  EXOr-type objects \citep{herbig89,herbig08}. In the last decade there has been an increased interest in the variability of { eruptive} young stars and its role in star formation \citep[see][for a review]{Audard:2014a}. Such { strong variability (of a few magnitudes)} at the earliest phases of star formation could explain   {\it the luminosity problem}  \citep{kenyon90}, i.e. the fact that the observed luminosity of young stars is too low compared to what is expected from the gas accretion rate that is needed to accumulate their mass within the required timeframe \citep{evans09,enoch09,dunham15}. 

It is expected that luminosity outbursts are more prominent during the initial phases of star formation Class 0, I objects) but only a few observations of varying YSOs have been reported \citep{kospal11,caratti11,Safron:2015a}. This may be due to (i) the short duration of the these early phases of star formation, or (ii) the short duration of an outburst episode (a few years to a few centuries). Furthermore, as young protostars are deeply embedded in their parent envelopes  the effect of an outburst may be difficult to observe, at least at optical wavelengths. Therefore, longer  wavelength, observations may be more appropriate  \citep[e.g][]{kospal07,hunter17,Safron:2015a,yoo17}. 

Recognizing the importance of longer wavelength observations, the James Clerk Maxwell Telescope (JCMT) TRANSIENT Survey is monitoring eight star forming regions within $500 \ \text{pc}$ \citep{herczeg17}, with the goal to observe variability at sub-mm wavelengths related to episodic accretion events. \cite{Johnstone:2018b} summarizes the results of the survey, which has observed significant variability in EC53 \citep{yoo17} and small changes on  other objects \citep{Mairs:2018a} 

 MacFarlane et al. (2019) (hereafter  Paper I), investigate the effect of
 episodic outbursts  on  the SED of a YSO self-consistently formed within a hydrodynamic simulation of a collapsing cloud, including episodic accretion leading to episodic luminosity outbursts. They performed polychromatic radiative transfer calculations using the properties (luminosity, density structure) of the young protostar as provided by the simulation. Here, in order to compare against a broader variety of outbursting young stars \citep{Audard:2014a},  we perform  an additional set of $25$ radiative transfer (RT) simulations, exploring a wider parameter space of outburst luminosities and YSO properties (disc mass, envelope mass, outflow cavity). 
 
This paper is structured as follows. In Section~\ref{sec:methods} we briefly describe the hydrodynamic simulation and the radiative transfer method used. In Section~\ref{sec:param_space} we discuss the effect of outbursts on the SED for different outburst luminosities and YSO structures. Finally, in Section~\ref{sec:conclusions} we summarize our results.}

\section{Computational method}
\label{sec:methods}
 
 We briefly describe the hydrodynamic simulation and the radiative transfer method. For a more detailed discussion, please see Paper I. We then discuss how we modify the properties of the YSO to cover a more extended parameter space.
 
 \subsection{Hydrodynamic simulation of YSO formation}
 
 We follow the evolution of a collapsing $5.4 \ \text{M}_\odot$  cloud  that forms an embedded protostar \citep{stamatellos12}, using the Smoothed Particle Hydrodynamics (SPH) code  SEREN \citep{hubber11a, hubber11b} with the radiative transfer method of \citet{stamatellos07}.  { If we assume a typical star formation efficiency of $\sim 20\%$ \citep[e.g.][]{Andre:2014a} then this $5.4 \ \text{M}_\odot$-cloud will end up forming roughly a solar-mass star}. The simulation employs a model to capture the episodic accretion of material onto the young protostar that results in episodic outbursts of luminosity \citep[for details see Paper I, ][]{Stamatellos:2011a, stamatellos12}, and in turn affecting the structure of the YSO \citep{MacFarlane:2017a}. The episodic behaviour is due to the effect of gravitational instability (GI)  operating in the outer disc, combined with the  magneto-rotational instability (MRI)  operating in the inner disc when the temperature is high enough to provide an appropriate level of ionisation  \citep{armitage01, zhu10a, zhu10b, Stamatellos:2011a, Mercer:2017a}.

 \subsection{Radiative transfer modelling}
 
We perform detailed polychromatic radiative transfer simulations using the 3D radiative transfer Monte Carlo code  RADMC-3D\footnote{\href{url}{http://www.ita.uni-heidelberg.de/$\sim$dullemond/software/radmc-3d/} (Last accessed: 07/07/2018)} \citep{dullemond12}. This code utilizes a  Ray-tracing Radiative Transfer (RRT), to produce synthetic observations (SEDs and images) that can then be compared with observations. To translate the SPH density profile to a grid \citep{Stamatellos:2005b} to be used by  RADMC-3D an adaptive-mesh refinement method is used \citep[][see Paper I]{Robitaille:2011a}. Our models take into account radiation from the young protostar (both intrinsic and due to accretion) and heating from the ambient interstellar radiation field (ISRF) \citep{andre03}. We assume a gas-to-dust ratio of 100, and we use the \citet{oh94} (OH5) opacities that refer to  a standard MRN mixture (\citealp{mrn77}; $53\%$ silicate, $47\%$ graphite) with thin ice mantles at a density of $10^{6}  \ \text{cm}^{-3}$.

\begin{figure}
   \centering
   \includegraphics[trim={-0.75cm, 0.75cm 0.0cm 0.0cm},width=0.99\columnwidth,keepaspectratio]{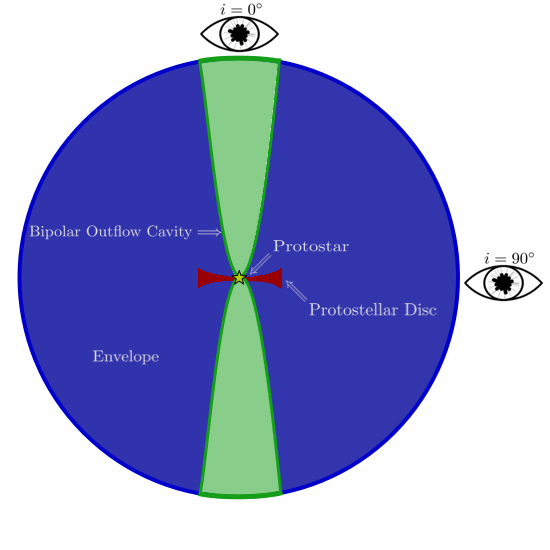}
\caption{Schematic representation of the different components of the modelled YSO: disc, envelope, and outflow cavity. The inclinations $i = 0^\circ \ \text{and} \ 90^\circ$ are also indicated.}
\label{fig:yso_schem}
\end{figure}

 \subsection{YSO model setup}

{ The parameters of the  RT models performed are presented in Table~\ref{tab:model_params}.   We chose an accretion event from the hydrodynamic simulation  (event E2; see Paper I) and we perform RT simulations on two snapshots: just before and near the end of the outburst event. The mass of the protostar at the two chosen snapshots are 0.26~M$_{\sun}$ and 0.42~M$_{\sun}$, respectively, and the corresponding luminosities a
 2.0~L$_{\sun}$ and 1110~M$_{\sun}$. {The high luminosity provided by the hydrodynamic simulation is due to an intense accretion event. Such events are thought to be rather common at the early phases of the formation of the protostar, and indeed  high outburst luminosities have been observed  e.g. Z CMa \cite[$400-600\text{L}_\odot$;][]{Hartmann:1996a} and  V1057 Cyg  \cite[$250-800\text{L}_\odot$;][]{Hartmann:1996a}.}
We artificially set the outburst luminosities to $10, \ 20, \ 50, \ \text{and} \ 80 \ \text{L}_{\odot}$, rather than $1110 \ \text{L}_{\odot}$ as directly provided by the hydrodynamic simulation (see Paper I).} { These luminosities have been chosen so as to correspond to the wide range of the luminosities of FU-Ori type outbursts observed \citep{Audard:2014a}, which cannot be captured by a single hydrodynamic simulation.} The quiescent phase protostellar luminosity is that of the simulation output, i.e. $L_{*,q} = 2 \ \text{L}_\odot$. The ratios of protostellar luminosity between the outbursting and the quiescent phase are therefore $5, \ 10, \ 25, \ \text{and} \ 40$, respectively.  { We therefore make the inherent assumption that the density profile of the YSO does not depend on the luminosity of the embedded protostar, which is not true \citep{MacFarlane:2017a}. However, our aim here is to investigate how radiation propagates (and eventually reaches the observer) through an asymmetric YSO, using a given density structure (provided by the hydrodynamic simulation). Moreover, the luminosity of the protostar affects mostly  the density structure near it whereas the large-scale structure remains mostly unaffected \citep[see][for details]{MacFarlane:2017a}.}
\begin{table}
\centering
	\caption{Parameters for the $25$  RT models. The outflow cavity opening angle at $r_{xy} = 10 \ 000 \ \text{AU}$ is set to $\theta_\text{c}=10^{\circ}$  for all models. Disc ($M_\text{d}$) and envelope ($M_\text{e}$) masses are given in solar masses and in  units of the corresponding masses from the simulation ($M_\text{d}^\text{s} \simeq 0.3 \ \text{M}_\odot$; $M_\text{e}^\text{s} \simeq 4 \ \text{M}_\odot$). The quiescent phase snapshot of  event E2 is used (where $L_{*} = 2 \ \text{L}_\odot$; see Paper I).}
	\renewcommand{\arraystretch}{1.3} 
	\begin{tabular}{ccccc}
		\hline
		 $M_\text{d}$/$M_{\sun}$ & $M_\text{d}/M_\text{d}^\text{s}$&   $M_\text{e}/M_{\sun}$& $M_\text{e}/M_\text{e}^\text{s}$  & $L_{*}$\\
		\hline
		  0.3&$1$ 	& 4	& $1$ 	& $2, \ 10, \ 20, \ 50, \ 80$ \\
		  0.03&$0.1$ 	& 4 & $1$ 	& $2, \ 10, \ 20, \ 50, \ 80$ \\
		  0.6&$2$ 	& 4	& $1$ 	& $2, \ 10, \ 20, \ 50, \ 80$ \\
		  0.3&$1$ 	& 0.4	& $0.1$ 	& $2, \ 10, \ 20, \ 50, \ 80$ \\
		  0.3&$1$ 	& 8	& $2$ 	& $2, \ 10, \ 20, \ 50, \ 80$ \\
		\hline
	\end{tabular}
\label{tab:model_params}
\end{table}

 In addition to varying the outburst luminosity, we also vary the mass of the different components of the YSO, to explore their impact on the SED during an episodic accretion event.   A schematic representation of the different components (disc, envelope and outflow cavity)  of the YSO considered  is presented in Fig.~\ref{fig:yso_schem}.  We perform RT simulations of a YSO for which the disc and envelope densities are each rescaled to $\times 0.1 \ \text{and} \ \times2$ times, respectively,  those of the simulation output. The disc-envelope radial boundary is set by the location at which the surface density falls below a value of $20 \ \text{g cm}^{-2}$. The vertical boundary of the disc is set to three times the disc scale height $h_\text{d}(r_{xy})$, where
\begin{equation}\label{eq:scale_height}
h_\text{d}(r_{xy})= \frac{c_{s}(r_{xy})}{\Omega_{K}(r_{xy})}.
\end{equation}
In the above formalism, $r_{xy}$ is the radius on the disc mid-plane, $c_{s}(r_{xy})$ is the sound speed, and $\Omega_{K}(r_{xy})$ is the Keplerian angular velocity.

Additionally, the model  accounts for bipolar outflows in the direction perpendicular to the disc midplane. We define the  surface  of the outflow cavity using %
\begin{equation}\label{eq:cavity}
|z(r_{xy})| =\alpha r_{xy}^{\beta},
\end{equation}
where $\alpha$ and $\beta$ are free parameters that determine the cavity opening angle $\theta_\text{c}$ and the cavity flaring
($\tan \theta_\text{c}=\alpha (r_{xy}^{_\text{c}})^{\beta-1}$), and $r_{xy}^{_\text{c}}$ is the radius on the disc midplane at which $\theta_\text{c}$ is calculated. { The density and geometry (i.e. the values of $\alpha$, $\beta$ above) of the outflow cavity depend on the evolutionary stage of the YSO. As we focus on Class~0 YSOs, motivated by the works of \citet{whitney03a, whitney03b}, we
set  $\beta = 3$, and the opening angle of the cavity to $\theta_\text{c}$=$10^{\circ}$ at $r_{xy}^{_{\text{c}}} = 10 \ 000 \ \text{AU}$. 
The cavity density is set to $n_{\text{H}_{2}} = 10^{5} \ \text{g cm}^{-3}$. At large distances from the star we ensure that the density of the cavity is not higher than the density of the adjacent envelope  by tapering this value.}

\section{Results}
\label{sec:param_space}

\subsection{Spectral Energy Distributions}

\begin{figure}
   \centering
   \includegraphics[trim={0.5cm 0.5cm 0.5cm 0.5cm},width=1.\columnwidth,keepaspectratio]{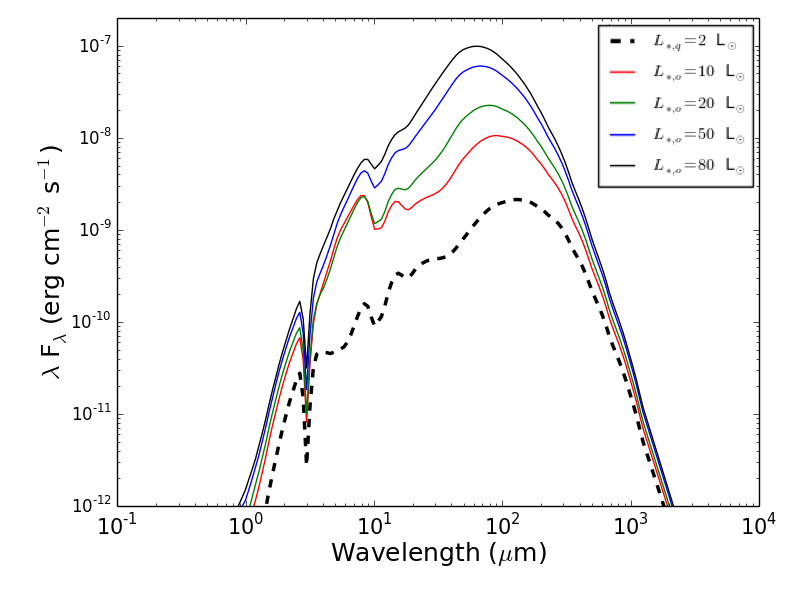}
\caption{SEDs for models with different protostellar luminosities, for unscaled YSOs at an inclination of $i=30^\circ$. The thick black dashed line corresponds to the quiescent phase  ($L_{*,q} = 2 \ \text{L}_\odot$). Solid red, green, blue and black lines represent outburst phases of the YSO, with protostellar luminosities of $L_{*,o} = 10, \ 20, \ 50, \text{and} \ 80 \ \text{L}_\odot$, respectively.}
\label{fig:seds_lumscale}
   \centering
   \includegraphics[trim={0.5cm 0.5cm 0.5cm 0.5cm},width=1.\columnwidth,keepaspectratio]{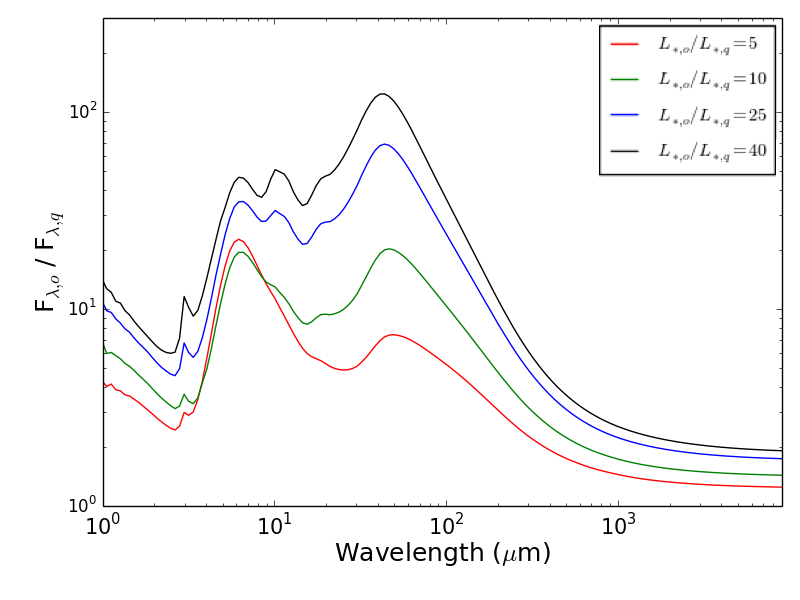}
\caption{The ratio of outburst to quiescent flux for the RT models presented in Fig.~\ref{fig:seds_lumscale}. The flux increase is more prominent in the NIR but still present at the FIR and submm/mm region.}
\label{fig:seds_lumscale_ratio}
\end{figure}

For each model, we compute SEDs for inclinations of $0^{\circ}, \ 30^{\circ}, \ 60^{\circ} \ \text{and} \ 90^{\circ}$, for radiation originating from $R \leq 10,000 \ \text{AU}$. { This is the typical size of a Class 0 object as mapped in the sub-mm (e.g. at 850~\micron\ by JCMT). In Section~\ref{sec:sed_area} we also also present RT models with integration areas $R \leq 1,000 \ \text{AU}$, corresponding to the central regions of YSOs.}
SEDs  are computed assuming a distance of $140 \ \text{pc}$. In Fig~\ref{fig:seds_lumscale} we present SEDs for RT models with unscaled disc and envelope components. SEDs are computed at an inclination of $i = 30^\circ$, and in Fig~\ref{fig:seds_lumscale_ratio} we present the ratio between the outburst and the quiescent flux. We see that the  maximum increase of flux occurs around 5-80~$\micron$ (a factor from $\sim10-100$ for a luminosity increase of a factor 5-40). However, there is also a detectable increase at longer wavelengths (e.g.  $\sim1.5-3$ at 850~$\micron$).

\begin{figure}
   \centering
   \includegraphics[trim={0.5cm 0.5cm 0.5cm 0.5cm},width=0.99\columnwidth,keepaspectratio]{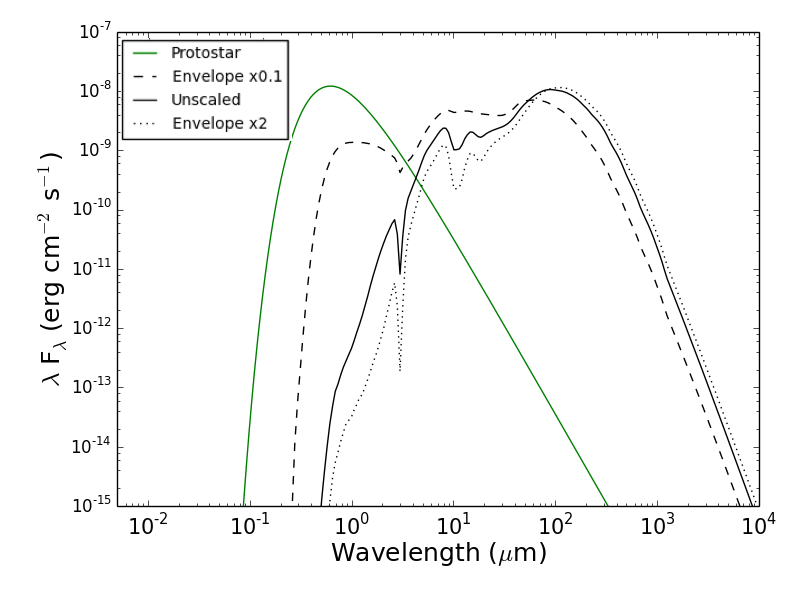}
\caption{SEDs of  YSOs with varying envelope mass for protostellar luminosity $L_{*,o} = 10 \ \text{L}_\odot$. Black dashed, solid, and dotted lines represent models where the envelope mass is scaled by $\times0.1$, $\times1$, and $\times2$, respectively. SEDs are computed  at an inclination of $30^{\circ}$. The green solid line corresponds to the protostellar emission.}
\label{fig:seds_params}
\end{figure}

In Fig.~\ref{fig:seds_params}, we present the SEDs for a representative model with an outbursting phase luminosity of $L_{*,o} = 10 \ \text{L}_\odot$, at an inclination of $i = 30^\circ$. Models with rescaled envelope mass ($\times0.1$, $\times1$ - unscaled, and $\times2$ the original value) are plotted. As expected, with decreased envelope mass the flux increases at short wavelengths { (below around $100-200~\micron$)} and decreases at longer wavelengths. Likewise, increasing the envelope mass results in  the reduction of short-wavelength flux and increase of flux at longer wavelengths. 

\begin{figure}
   \centering
   \includegraphics[trim={0.5cm 0.5cm 0.5cm 0.5cm},width=0.99\columnwidth,keepaspectratio]{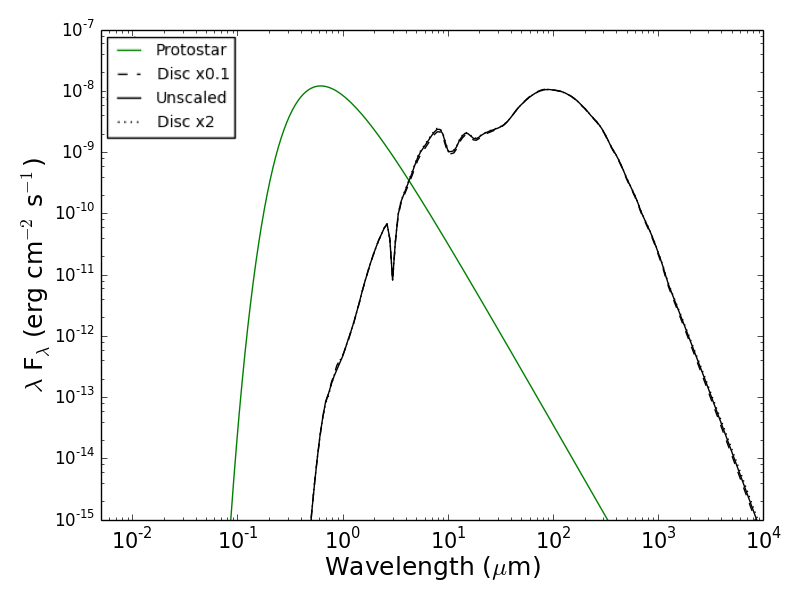}
\caption{As per Fig.~\ref{fig:seds_params}, but for varying disc mass. Black dashed, solid and dotted lines correspond to models with disc masses scaled to $\times0.1$, $\times1$ and $\times2$ the hydrodynamic simulation value, respectively.}
\label{fig:seds_params2}
\end{figure}

We also present SEDs comparing models with rescaled disc mass in Fig.~\ref{fig:seds_params2}. The impact of disc mass rescaling is minimal when compared to the  envelope-rescaled models. This is because the rescaled disc mass is of order of $0.3 \ \text{M}_\odot$, whereas the envelope mass is of order of $4 \ \text{M}_\odot$. The change in the total mass is therefore much larger when the mass of the envelope is scaled. It is expected that the effect of the disc will be more prominent when viewing the disc edge-on.


\subsection{Flux increases during outbursts}

\begin{figure*}
   \centering
   \subfigure{\includegraphics[trim={0.5cm 0.5cm 0.5cm 0.3cm},clip,width=0.95\columnwidth,keepaspectratio]{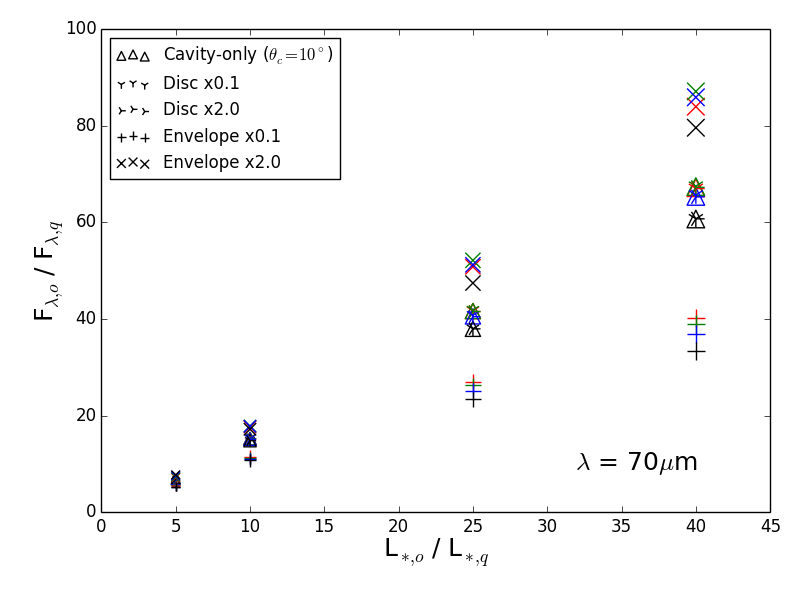}}
   \subfigure{\includegraphics[trim={0.5cm 0.5cm 0.5cm 0.3cm},clip,width=0.95\columnwidth,keepaspectratio]{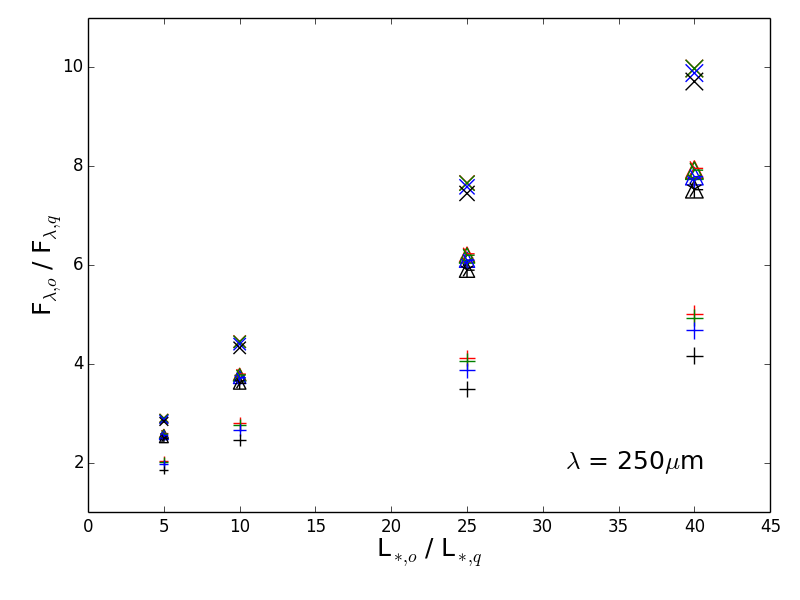}}
   \subfigure{\includegraphics[trim={0.5cm 0.5cm 0.5cm 0.3cm},clip,width=0.95\columnwidth,keepaspectratio]{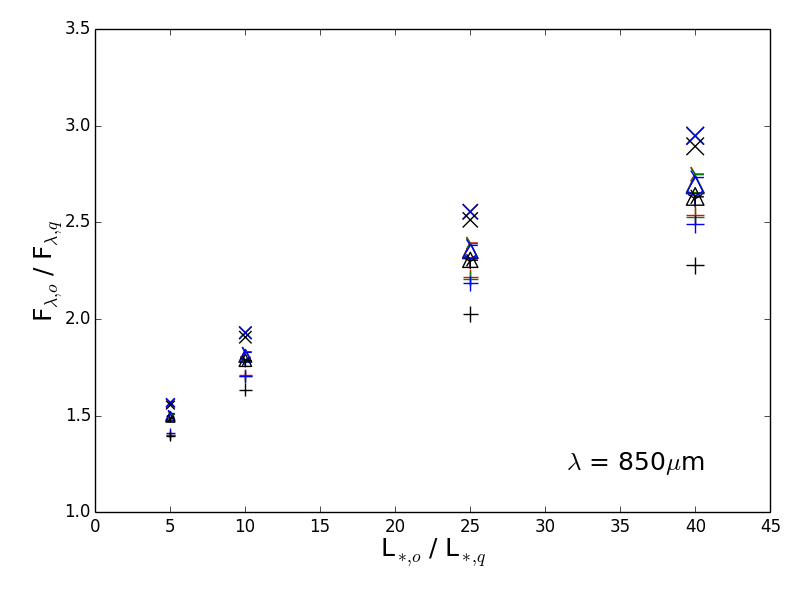}}
   \subfigure{\includegraphics[trim={0.5cm 0.5cm 0.5cm 0.3cm},clip,width=0.95\columnwidth,keepaspectratio]{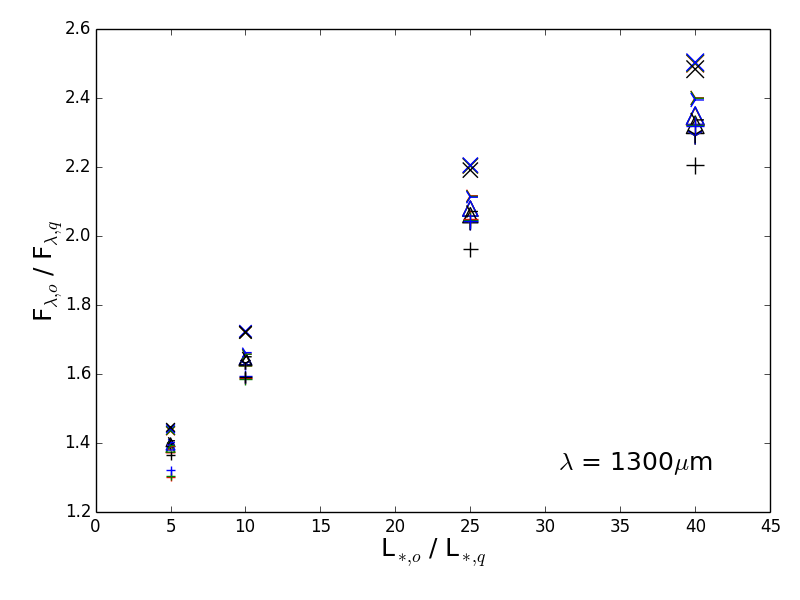}}
\caption{Outburst-to-quiescent flux ratios, $F_{\lambda, o} / F_{\lambda, q}$, at $70, \ 250, \ 850, \ \text{and} \ 1300 \ \mu\text{m}$,  as a function of the protostellar luminosity $L_{*, o}$,  computed  integrating over the inner $R \leq10,000$~AU. Red, green, blue and black markers correspond to SEDs at $0^{\circ}, \ 30^{\circ}, \ 60^{\circ}, \  \text{and} \ 90^{\circ}$, respectively. Triangles represent the default model (no rescaling),  upward/downward  tri-spoke markers are  models with disc masses rescaled to 0.1/2 times the simulation values, and plus/cross markers are models with envelope masses scaled 0.1/2 times.}
\label{fig:ratio_params}
\end{figure*}

For each outburst phase protostellar luminosity, we compute the outbursting-to-quiescent flux ratios at $70, \ 250,  \ 850 \ \text{and} \ 1300 \ \mu\text{m}$, i.e.  wavelengths corresponding to continuum bands of the {\it Herschel Space Observatory}, JCMT and Atacama Large (Sub-)Millimetre Array (ALMA). By computing the flux response due to a known increase in protostellar luminosity, we provide a diagnostic tool for future observations to determine the protostellar luminosity increases responsible for increases to long wavelength emission. 

Flux ratios are plotted for all models in Fig.~\ref{fig:ratio_params}, for $70 \ \mu\text{m}$ (top left panel), $250 \ \mu\text{m}$ (top right panel), $850 \ \mu\text{m}$ (bottom left panel) and $1300 \ \mu\text{m}$ (bottom right panel).
We find that the flux ratio increases at all wavelengths with increasing ratio of outbursting-to-quiescent protostellar luminosity, but that the flux increase is smaller at longer wavelengths. Illustrating this further, we refer to Figs.~\ref{fig:seds_lumscale}-\ref{fig:seds_lumscale_ratio}, where we see that between the quiescent phase (dashed line) and outburst phase (solid lines) SEDs, the greatest increase in flux occurs around  5-80~$\micron$.  For the flux ratios at $70$ and $250 \ \mu\text{m}$, the response to the protostellar luminosity increase is not significantly different for models with different disc masses. On the other hand, higher envelope masses result in higher flux ratios. The contribution from the disc is rather small, considering that it has much smaller mass than the envelope itself. For the flux ratios at $850$ and $1300 \ \mu\text{m}$, there are only small differences between different models. 

\subsection{YSO classification}

\begin{figure}
   \centering
   \includegraphics[trim={0.5cm 0.5cm 0.5cm 0.5cm},width=0.9\columnwidth,keepaspectratio]{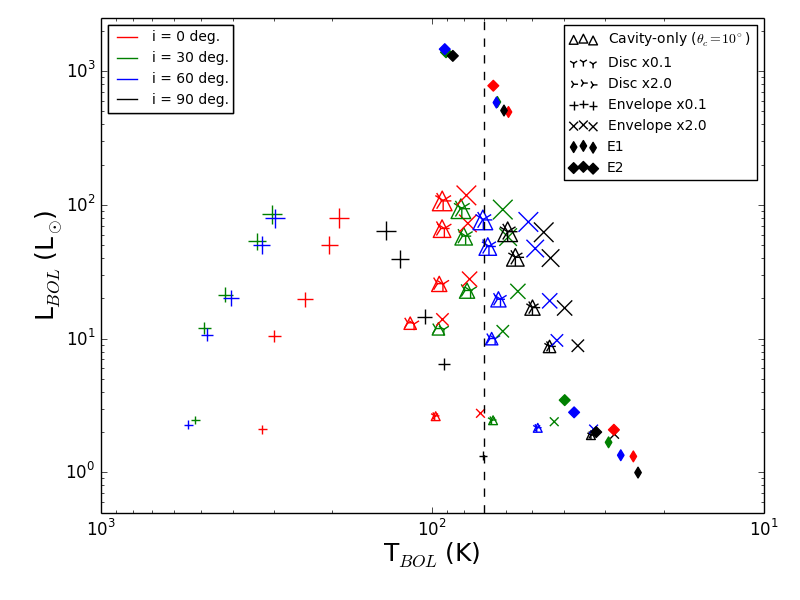}
\caption{$L_\text{BOL}$ as a function of $T_\text{BOL}$. Markers and colours as per Fig.~\ref{fig:ratio_params}. Diamond markers denote data from models adopting quiescent/outbursting protostellar luminosities from the hydrodynamic simulation, where thin/thick markers are for E1/E2 events. Symbol sizes are proportional to the luminosity of the protostar. The dashed vertical line represents the Class 0/I YSO boundary.}
\label{fig:lbol_tbol}
\end{figure}

We compute bolometric values for the different YSOs and in Fig.~\ref{fig:lbol_tbol} we plot the bolometric luminosity $L_\text{BOL}$ against  the bolometric temperature $T_\text{BOL}$ \citep{myersladd93}. Our aim is to consider the impact of the varying protostellar luminosities and different masses of  YSO components on the calculated values (and the subsequent classification of the YSO).  We also plot the values derived from the YSO in the hydrodynamic simulation (see Paper I for description of the two events E1 and E2) (thin diamonds: E1; thick diamonds: E2). The vertical dashed line in Fig.~\ref{fig:lbol_tbol} indicates the $T_\text{BOL}$ boundary of $70 \ \text{K}$ between YSO Class 0 and I \citep{chen95}.  { The bolometric temperature generally increases with decreasing inclination, i.e. when looking through the outflow cavity, as the hotter regions close to the protostar are probed. The only exception is  when the mass of the envelope is scaled down by a factor of 10; this is due to the envelope asymmetric structure which is more pronounced relative to the outflow cavity (in this model the density of the cavity is not much different from the density of envelope itself;  also  see the discussion on envelope asymmetries in Paper I).}

\begin{figure*}
   \centering
   \subfigure{\includegraphics[trim={0.5cm 0.5cm 0.5cm 0.3cm},clip,width=0.95\columnwidth,keepaspectratio]{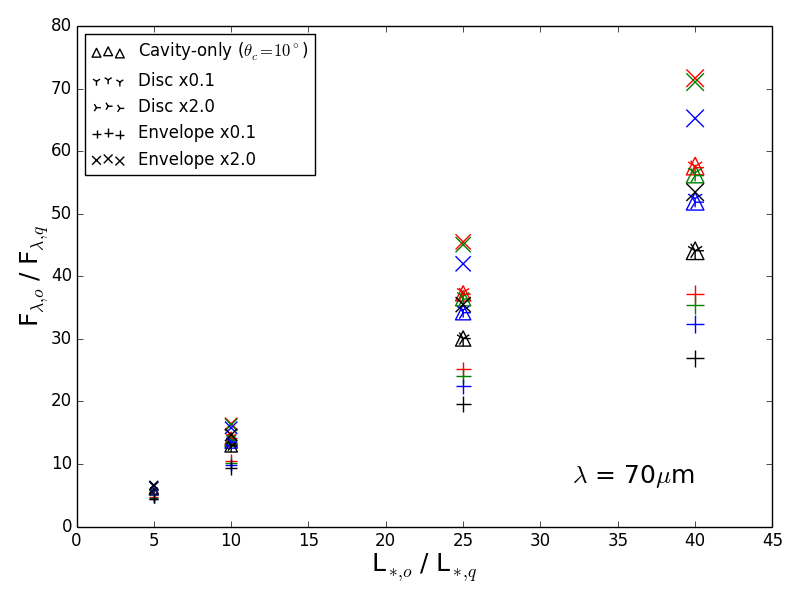}}
   \subfigure{\includegraphics[trim={0.5cm 0.5cm 0.5cm 0.3cm},clip,width=0.95\columnwidth,keepaspectratio]{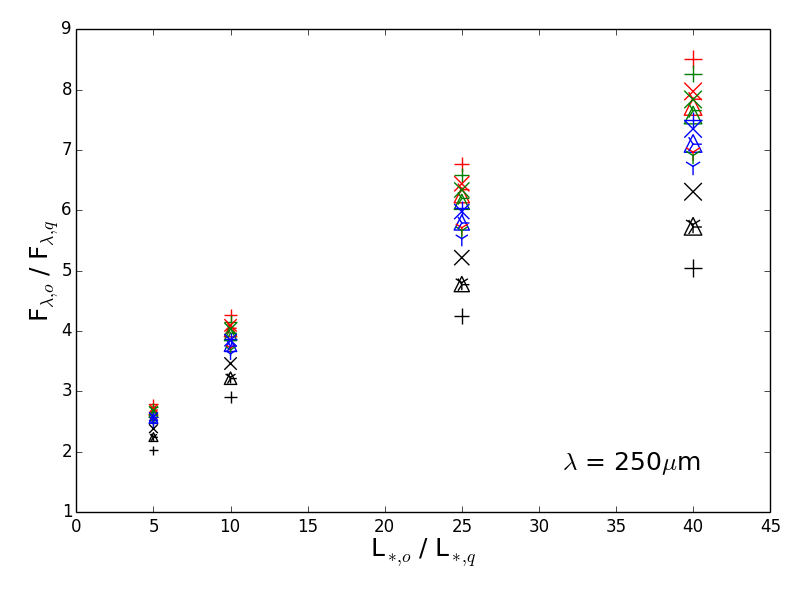}}
   \subfigure{\includegraphics[trim={0.5cm 0.5cm 0.5cm 0.3cm},clip,width=0.95\columnwidth,keepaspectratio]{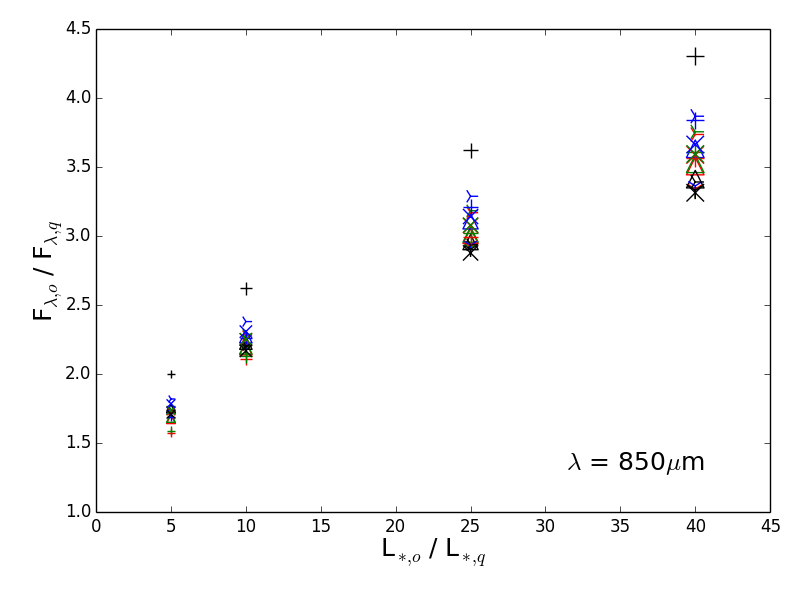}}
   \subfigure{\includegraphics[trim={0.5cm 0.5cm 0.5cm 0.3cm},clip,width=0.95\columnwidth,keepaspectratio]{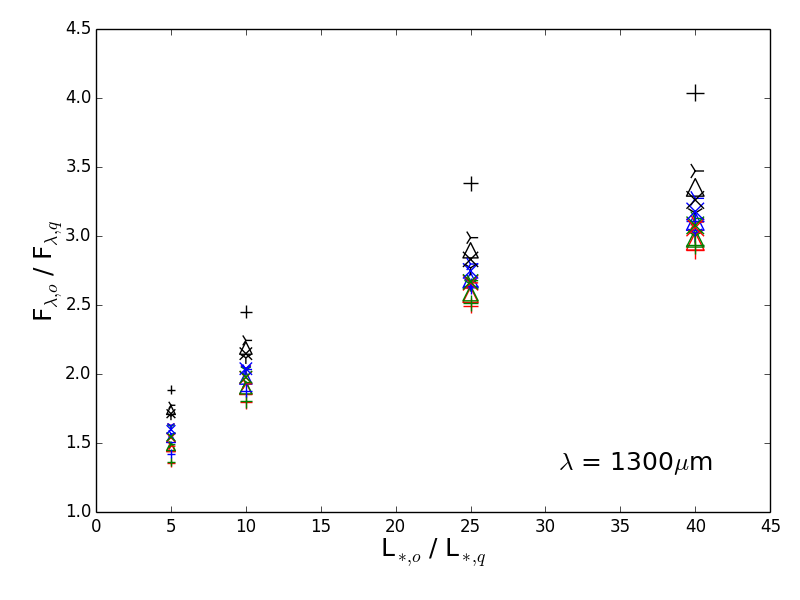}}
\caption{Outburst-to-quiescent flux ratios, $F_{\lambda, o} / F_{\lambda, q}$, at $70, \ 250, \ 850, \ \text{and} \ 1300 \ \mu\text{m}$, as a function of the outburst protostellar luminosity $L_{*, o}$,  computed  integrating over the inner $R \leq1,000$~AU (symbols are the same as in Figure~\ref{fig:ratio_params}).}
\label{fig:ratio_params_1e3}
\end{figure*}

All the different YSOs that we model here (Fig.~\ref{fig:lbol_tbol}) are by definition Class 0 objects as there is more mass in the envelope than in the protostar itself. However, we note that the models in which we decrease the envelope mass by a factor of 10 are on the boundary between Class~0 and Class~I phase ($M_*\approx M_e$). The observed bolometric temperature is found to be sensitive to the YSO inclination, and it may change the classification from Class~0 to Class~I if a strict value of 70~K is used. The influence of geometry on bolometric properties folding into YSO classification has been previously noted (e.g.  \citealp{dunham12,dunham13}{and Paper I}). If we use a boundary value of 100~K instead of 70~K then most of the YSOs would be appropriately classified as Class 0 objects, with the exception of the low envelope-mass models which exhibit high bolometric temperatures and would be still classified as Class I objects (but nevertheless they are in the boundary between the two phases). { However,  the definition of a strict boundary is not possible without exploring a wider parameter space as there is a dependence on the YSO luminosity and orientation.}

\begin{figure}
   \centering
   \includegraphics[trim={0.5cm 0.5cm 0.5cm 0.5cm},width=0.9\columnwidth,keepaspectratio]{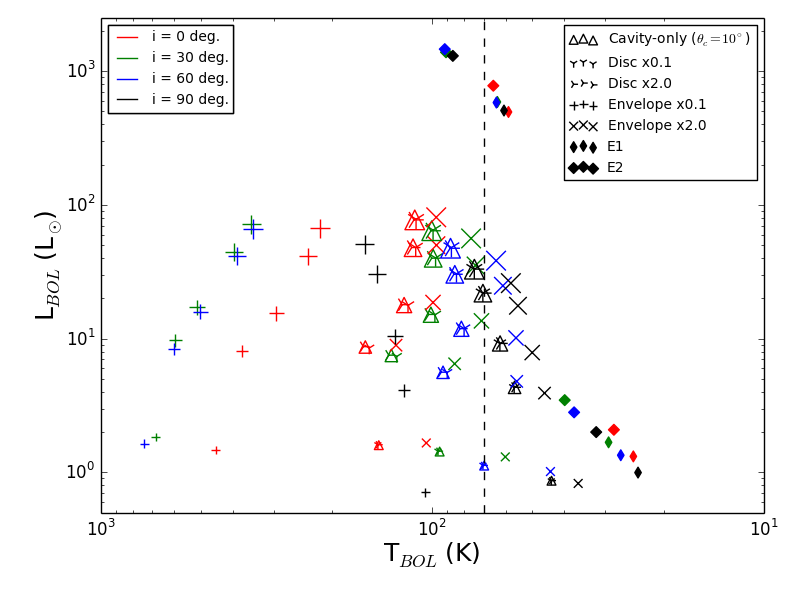}
\caption{$L_\text{BOL}$ as a function of $T_\text{BOL}$ for the RT models  in Table~\ref{tab:model_params}. Markers and colours as per Fig.~\ref{fig:lbol_tbol}. Figure as per Fig.~\ref{fig:lbol_tbol}, but the bolometric properties are computed including flux only from the inner $R \leq 1000 \ \text{AU}$ region of the YSO.}
\label{fig:lbol_tbol_1e3}
\end{figure}

\subsection{Impact of the integration area}
\label{sec:sed_area}

{ High resolution observations (e.g. by ALMA) can probe the very central regions of YSOs, providing better constraints on the properties of embedded young protostars. We perform additional RT simulations, computing SEDs integrating over the inner $R \leq 1,000 \ \text{AU}$ region of the YSO, which corresponds to an angular size of $2\arcsec-7\arcsec$  at typical distances of nearby star forming regions (500 to 140~pc).} We then compute the flux ratios and discuss the corresponding $L_\text{BOL} - T_\text{BOL}$ diagnostics, to understand the impact of the integration area on the  flux changes during outbursts, and on the YSO classification.

\subsubsection{Flux ratio changes}

We present the flux ratios between the quiescent and outbursting phases, at wavelengths of 70,  250, 850, and 1300~\micron\ in Fig.~\ref{fig:ratio_params_1e3}. We find a similar behaviour to the previous analysis  (i.e.  integrating over the inner $R \leq 10,000 \ \text{AU}$ region), i.e. all flux ratios increase with increasing outburst protostellar luminosity.

At $70 \text{ and } 250 \ \mu\text{m}$, we find that the flux ratios are smaller  in comparison with those computed over a more extended region ($R \leq 10,000 \ \text{AU}$). They also exhibit smaller differences between models. With decreasing integration area over which the flux ratios is computed, the effect of varying the envelope mass becomes unimportant, as for the smaller region considered, the YSO mass is dominated by the disc mass. As the changes in the disc mass are relatively small between different models. It follows that for all models, more centrally located regions have increasingly similar masses, and therefore similar continuum fluxes.

Both the $850 \text{ and } 1300 \ \mu\text{m}$ continuum wavelengths exhibit higher flux ratio increases for SEDs computed over smaller spatial scales with the YSO (compare Figure~\ref{fig:ratio_params} with Figure~\ref{fig:ratio_params_1e3}). This is a result of not including the outer regions of the envelope, which are primarily heated by the ISRF. The inclusion of the outer YSO regions attenuates the effect of the protostellar luminosity on the SEDs.  The result shown here serves as clear motivation to pursue high resolution monitoring of episodically accreting protostars, particularly at (sub-) mm wavelengths, to ensure sufficient contrast in flux between quiescent and outbursting phases, whilst at the same time minimising the effect of the ISRF.

\subsubsection{YSO classification for a smaller integration area}
\label{sec:classification}
We present $L_\text{BOL}$ and  $T_\text{BOL}$ values for the SED integrated over the $R \leq 1000 \ \text{AU}$ region in Fig.~\ref{fig:lbol_tbol_1e3}.  Two clear trends are found when comparing with the results computed considering a larger integration area ($R \leq 10000 \ \text{AU}$, see Fig.~\ref{fig:lbol_tbol}): $T_\text{BOL}$ increases,  as we consider only the hotter inner region, while  $L_\text{BOL}$ decreases, as we do not take into account a significant portion of the radiation from the extended envelope.
We find that by proxy of the $T_\text{BOL} = 70 \ \text{K}$ boundary, all YSOs models in which the envelope mass is reduced  are classified as Class I objects. Unscaled, and disc-mass-scaled models at inclinations lower than $90^\circ$ also correspond to Class I YSOs. In contrast, the most heavily embedded (with twice as much mass in the envelope), and lowest luminosity, edge-on models with no scaling, or decreased/increased disc mass, are classified as Class 0. This result confirms that the SED integration area also affects the bolometric properties of YSOs.

\section{Conclusions}\label{sec:conclusions}

We have explored a range of radiative transfer models of episodically outbursting YSOs with a variety of structure and luminosities, with the aim to provide a diagnostic study to estimate the change in luminosity of episodically outbursting embedded protostars using long wavelength observations.  

We based our YSO models on the results of hydrodynamic simulations that include episodic feedback (see Paper I) but  we extended the YSO parameter space by varying the protostellar luminosity and the masses of the different components of the YSO (disc, envelope, outflow cavity) to  compute SEDs over the central $R \leq 10,000 \ \text{AU}$ of the YSO at different inclinations.  An additional set of SEDs were computed, integrating flux over the central $R \leq 1,000 \ \text{AU}$ region of the YSO. 

We calculated the flux increase at different wavelengths during episodic accretion events, and we found a reduction in outburst-to-quiescent flux ratio with increasing wavelength. Long wavelength emission shows only a small flux ratio increase during an outburst (an increase by a factor of 1.3 to 2.6  at 1.3mm). In contrast, there are more significant differences in the flux  ratio  at shorter (far-IR) wavelengths (an increase by a factor of 10 to 90 at 70\micron). { This is similar to what is seen for EC53 \citep{yoo17}, and  also for the outbursting protostar 2MASS 22352345+7517076 that has been observed at many different wavelengths by various telescopes \citep{Kun:2019a}.}

When the flux is computed over the $R \leq 1000 \ \text{AU}$ region, it was found that there is only a moderate increase of the flux ratio at far-IR wavelengths (an increase by a factor of 5 to 70  at 70~\micron). { At (sub-)mm wavelengths when integrating  over a smaller region around the  YSO, the flux  ratio increase is larger than when integrating over an area of 10,000~AU (an increase by a factor of 1.5  to 4  at 1.3mm). This provides evidence that high resolution observations \citep[e.g. with ALMA, see][] {Francis:2019a} can be used to provide more reliable estimates of the change of the intrinsic luminosity of protostars undergoing episodic accretion. }

We calculated the bolometric luminosity and temperature of the YSOs of different models  and we found that the bolometric temperature is  sensitive to the YSO inclination, and it may affect the classification of a YSO if a strict value of 70~K is used to distinguish between Class 0 and Class I objects. If we use a boundary value of 100~K instead of 70~K we obtain a more accurate classification.

The work presented in this paper describes the observational characteristics of young Class 0/I protostars undergoing episodic phases of mass accretion and therefore luminosity outbursts, akin to FU Ori-type objects. Our results provide diagnostics to infer the  luminosity of episodically outbursting embedded protostars using observations at FIR and mm wavelengths.

\section*{Acknowledgements}
BM is supported by STFC grant ST/N504014/1. DS is partly supported by STFC grant ST/M000877/1. DJ is supported by the National Research Council Canada and by an NSERC Discovery Grant.
GH is supported by general grant 11773002 awarded by the National Science Foundation of China. Simulations were performed using the UCLAN HPC facility and the COSMOS Shared Memory system at DAMTP, University of Cambridge operated on behalf of the STFC DiRAC HPC Facility. This equipment is funded by BIS National E-infrastructure capital grant ST/J005673/1 and STFC grants ST/H008586/1, ST/K00333X/1. {\sc Seren} has been developed and maintained by David Hubber, who we thank for his help. 
Column density maps were generated using the visualization software SPLASH \citep{price07}. This work is supported by the JCMT-Transient Team.
%


\bibliographystyle{mnras}

\bibliography{macfarlane18ii} 



\bsp	
\label{lastpage}
\end{document}